\documentclass[amsmath,amssymb,pre,twocolumn,superscriptaddress,nofootinbib]{revtex4-2}

\usepackage{latexsym}
\usepackage{amssymb}
\usepackage{mathrsfs}
\usepackage{amsmath}
\usepackage{soul}
\usepackage{bm}
\usepackage{verbatim}
\usepackage{epsfig}
\usepackage{multirow}
\usepackage{xspace}
\usepackage{ltablex}
\usepackage{ulem}
\usepackage{float}
\usepackage[english]{babel}
\usepackage[utf8x]{inputenc}
\usepackage[T1]{fontenc}
\usepackage[usenames,dvipsnames]{xcolor}

\usepackage{color}
\definecolor{darkblue}{rgb}{0,0,0.6}
\definecolor{darkred}{rgb}{0.6,0,0}

\usepackage{hyperref}
\hypersetup{
bookmarksopen=true,
pdftitle="",
pdfauthor="",
pdfsubject="",
pdfstartview={FitH},
pdfmenubar=true,
pdfhighlight=/O,
colorlinks=true,
urlcolor=darkblue,
citecolor=darkblue,
linkcolor=MidnightBlue,
}

\usepackage[capitalise]{cleveref}

\begin{document}

\title{Mean-field description for the architecture of low-energy excitations in glasses}

\author{Wencheng~Ji}
\author{Tom~W.J.~de~Geus}
\author{Elisabeth~Agoritsas}
\author{Matthieu~Wyart}

\affiliation{
    Institute of Physics,
    \'Ecole Polytechnique F\'ed\'erale de Lausanne (EPFL),
    CH-1015 Lausanne,
    Switzerland}

\date{\today}

\renewcommand{\floatpagefraction}{.3}

\begin{abstract}

    In amorphous materials, groups of particles can rearrange locally into a new stable configuration.
    Such elementary excitations are key as they determine the response to external stresses, as well as
    to thermal and quantum fluctuations.
    Yet, understanding what controls their geometry remains a challenge.
    Here we build a scaling description of the geometry and energy of low-energy excitations in terms
    of the distance to an instability, as predicted for instance at the dynamical transition in
    mean field approaches of supercooled liquids.
    We successfully test our
    predictions in ultrastable computer glasses, with a gapped and ungapped (regular) spectrum.
    Overall, our approach explains why excitations become less extended, with a higher energy and
    displacement scale upon cooling.

\end{abstract}

\maketitle

\section{Introduction}

If a liquid is cooled rapidly enough to avoid crystallization,
its dynamics rapidly slows down until the glass transition where equilibration cannot be achieved:
a glass is formed, and the material acts as a solid \cite{ediger1996supercooled}.
What controls the dynamics in such supercooled liquids is a long-standing question of condensed
matter \cite{Cavagna09,arceri_landes_2020_Arxiv-2006.09725}.
Yet, new observations further constrain the descriptions of this phenomenon.
The `swap' Monte-Carlo algorithms \cite{Glandt84}
(in which nearby poly-disperse particles can exchange positions, in addition to their usual
translation move)
can speed up the dynamics by 15 orders of magnitude or more,
and can change the glass transition temperature $T_g$ by up to a factor two \cite{Ninarello17}.
Because swap algorithms achieve thermal equilibrium,
theories of the glass transition in which thermodynamics governs kinetics \cite{Lubchenko01,Adam65}
appear ill suited to explain such a dramatic difference \cite{Wyart17}
(see \cite{Berthier19} for an alternative view).
Several theoretical works (including real-space \cite{Brito18}, replica \cite{Ikeda17} and
mode-coupling \cite{szamel2018theory} approaches)
predict that the dynamical transition temperature $T_c$ below which thermal activation becomes the
dominant mechanism of relaxation \cite{Cavagna09} decreases in the presence of swap,
plausibly explaining the speed up of this algorithm.
However, understanding the dynamics in the vicinity of $T_c$ in finite dimension $d$ remains a
challenge.
By contrast, in the infinite dimensional limit \cite{book_parisi_urbani_zamponi_2020}, mean-field
treatments are exact:
one finds that for ${T<T_c}$ a gap appears in the vibrational spectrum such that there are no
vibrational modes with a frequency $\omega < \omega_c$, whereby $\omega_c$ grows upon cooling
\cite{Franz15},
and that the relaxation time diverges at ${T\rightarrow T_c^+}$ \cite{Maimbourg16}.
For finite $d$, the vibrational spectrum instead presents a pseudo-gap (\textit{i.e.}\ the spectrum
vanishes as a power-law for small $\omega$) consisting of quasi-localized modes (QLMs)
\cite{Schober93}.
Moreover, thermally activated events or `hopping processes' still occur for ${T<T_c}$, leading to a
finite relaxation time.
What controls their architecture and energy scale is unclear.

In a parallel development,
there has been recently a considerable effort to analyze both QLMs as well as elementary
excitations (minimal rearrangements leading to a new metastable state, which form the building
blocks of hopping processes) as a function of glass stability
\cite{Baity15,Lerner16,Mizuno17,Lerner18,Shimada18,Scalliet19,Wang19,Khomenko20,Rainone20}.
Numerically, liquids are equilibrated at a parent temperature $T_p$ before being rapidly quenched
to ${T=0}$,
thus obtaining an inherent structure where the Hessian of the energy can be analyzed,
and where excitations can be triggered using a short thermal cycle.
Strikingly, it is found that the density of excitations is reduced by several decades as $T_p$
decreases \cite{Khomenko20},
and that the characteristic number of particles involved rapidly decreases upon cooling.
The former observation is consistent with recent experiments on vapor-deposited glasses
\cite{Liu14,Perez14}.
Both facts are unexplained.

In this article,
\textit{(i)}~we use mean-field and real-space arguments to express the typical scales, namely the
length $\ell_{\text{loc}}$,
displacement $\delta_{\text{loc}}$,
number of particles $N_{\text{loc}}$
and energy $E_{\text{loc}}$ of low-energy excitations,
assuming the presence of an underlying gap of magnitude $\omega_c$ in the vibrational spectrum.
We find our predictions to be accurately satisfied in gapped glasses \cite{Kapteijns19,Wencheng20}.
\textit{(ii)}~Our analysis implies scaling relations between local properties
$\ell_{\text{loc}},\delta_{\text{loc}},N_{\text{loc}},E_{\text{loc}}$,
that we find to be also satisfied also in \textit{regular} ultrastable (gapless) glasses.
These predictions give a new handle to study the relaxation of glasses.
Together with mean field result predicting a growing gap $\omega_c$ upon cooling, they also
explain why low-energy excitations become smaller with a higher energy as the glass stability
increases.

The outline of the paper is as follows.
In \cref{sec-scaling-description}, we provide a scaling description for typical features of local
excitations.
We then test its predictions first on gapped glasses in \cref{sec-gapped-glasses}, and second on
regular ultrastable glasses in \cref{sec-regular-glasses}.
We conclude in \cref{sec-conclusion}.

\section{Scaling description for local excitations}
\label{sec-scaling-description}

We construct scaling relations for local excitations' typical volume, length, and particle
displacement as a function of an underlying $\omega_c$.
Since the dynamic transition in a mean-field description of liquids corresponds to the point where
the Hessian of the energy becomes stable \cite{Lubchenko07,Franz15}, we consider a material with a
stability control parameter $\epsilon>0$ \cite{DeGiuli14}.
An instability driven by temperature occurs when $T$ approaches $T_c^-$, and we thus consider
${\epsilon \sim (T_c-T)}$.
Infinite dimensional \cite{Lubchenko07,Franz15} calculations as well as effective medium theory
\cite{DeGiuli14} then predict a vanishing minimal eigenvalue $\lambda_c$ of the Hessian that
generically depend linearly on $\epsilon$, corresponding to a gap frequency
${\omega_c \sim \sqrt{\lambda_c}\sim \sqrt{\epsilon}}$
above which the spectrum of the Hessian of the energy is a semicircle.
In finite dimensions, hopping processes between stable configurations will, however, occur.

To estimate the hopping processes' spatial extension, we consider two replicas of the system in the
glass phase $\epsilon>0$,
and denote by $Q(r)$ their overlap.
$Q(r)$ characterizes the similarity between two configurations at location $r$,
and is unity if they are identical
\footnote{
    The overlap between two configurations $A$ and $B$ can be defined for instance from the square
    of their Euclidean distance ${Q_{AB}=\exp(-\Delta_{AB})}$ where
    ${\Delta_{AB}=\frac{1}{N} \sum_{i=1}^N (\vec{x}_i^A - \vec{x}_i^B)^2 }$.
    A local overlap $Q_{AB}(r)$ is then easily obtained by restricting this sum to particles close
    to position $r$, see also \cite{franz_2012_PNAS109_18725,guiselin2020JChemPhys}.
}.
In (infinite-dimensional) mean field, the free energy of this coupled system undergoes a
saddle-node bifurcation as ${\epsilon\rightarrow 0}$ \cite{franz1998effective} at which point the
overlap is finite; we denote its value $Q^*$.
To describe the spatial fluctuations of $Q(r)$, we use the following Ginzburg-Landau free energy
\cite{biroli2018random,franz2011field} \footnote{In \cite{franz2011field}, it is shown that
    an additional term should enter \cref{1}.
    This relevant term is equivalent to spatial fluctuations of $\epsilon$, and is also present in
    descriptions of the Random Field Ising Model.
    As discussed below, the success of our approach suggests that this term is small in our glasses,
    and will affect physical properties only on length scales beyond those we can reach.
}, that schematically reads:
\begin{equation}
    \label{1}
    F[Q]
    = \int d^dr \left[- \epsilon (Q-Q^*)+ \frac{1}{3} (Q-Q^*)^3 + \frac{1}{2} (\nabla Q)^2 \right]
\end{equation}
At a finite $\epsilon$, this free energy has a local minimum at an homogeneous overlap
$Q_{\text{eq}}$ such that ${Q_{\text{eq}}-Q^*=\sqrt{\epsilon}}$, which characterizes the `distance'
to the instability.
By performing an expansion around $Q_{\text{eq}}$, we obtain for overlaps close to the local
minimum (only keeping terms that depend on $Q$):
\begin{equation}
    F[Q] \approx \int d^dr \left[ \sqrt{\epsilon}( Q-Q_{\text{eq}})^2 +\frac{1}{2} (\nabla Q)^2 \right]
    \, .
\end{equation}
The overlap $Q(r)$ displays thermal fluctuations, whose length scale and correlation volume can be
deduced from the correlation function $G(r)=\langle
    (Q(r)-Q_{\text{eq}})(Q(0)-Q_{\text{eq}})\rangle$.
For the quadratic free energy of \cref{1}, when $r/\xi<1$ with $\xi$ the correlation length, this
classical computation (see \cref{sec-derivation-correlation-function})) gives:
\begin{equation}
    \label{2}
    G(r)\sim \frac{1}{r^{d-2}} \exp(-r/\xi) \quad \text{with } \ \xi\sim\epsilon^{-1/4}
    \, .
\end{equation}
A similar length scale was predicted to affect the dynamics in mode-coupling theory
\cite{franz2000non,Biroli06}
and was observed to characterize the linear response near an instability \cite{Lerner14}.
\cref{2} also leads to a characteristic volume in which fluctuations are correlated.
The typical volume $V$ of an excitation can be related to the correlation length $\xi$ by the
spatial integration
${V \sim \int \text{d}^d r \, G(r) \sim \xi^2 \int \text{d} \tilde{r} \, \tilde{r} \,
            \exp{(-\tilde{r})} \sim \xi^2}$
(where $\tilde{r} \equiv r / \xi$), which is independent of dimension.
Such a quadratic relation between volume and length is already known to hold near jamming
\cite{Yan16,Shimada18}.
We thus have ${V \sim \xi^2\sim 1/\sqrt{\epsilon}}$.
In $d=3$, it implies dimensionally that $V \sim d_0 \xi^2$ , where $d_0$ is the characteristic
particle size.
In what follows we make the natural assumption that low-energy elementary excitations do occur on
the characteristic volume and length scale of spontaneous fluctuations,
such that their number of particles ${N_{\text{loc}}\sim V}$
and their length ${\ell_{\text{loc}}\sim \xi}$.

To obtain the characteristic displacement and energy scale of such local excitations,
we perform an expansion of a symmetric double well
${E(X) = - m \omega_c^2 X^2 + \chi X^4 +o(X^4)\equiv E_2 + E_4+o(X^4)}$ (for an asymmetric double
well, both the energy barrier and difference generically scale as the result we obtain
\cite{Wencheng20}).
Here ${X}$ is the norm of the displacement field of the excitation,
that satisfies ${X^2 \sim N_{\text{loc}} \delta_{\text{loc}}^2}$
where $\delta_{\text{loc}}$ is the typical particle displacement.
By analyzing the extrema of $E(X)$, one readily obtains that the energy barrier between the two
minima is ${E_{\text{loc}}\sim m^2\omega_c^4/\chi}$ \footnote{
    In \cite{Wencheng20}, $\chi$ was assumed to be a constant, leading to the incorrect relation
    ${E_{\text{loc}} \sim \omega_c^4}$.
}
and the distance between the local minima follows ${X^2\sim m\omega_c^2/\chi}$,
implying that ${\delta_{\text{loc}}^2\sim m \omega_c^2/(\chi N_{\text{loc}})}$.

Ultimately, the term ${E_4\equiv\chi X^4}$ stems from the quartic non-linearity in the
inter-particle interaction potential (which we assume to be short ranged).
We denote its characteristic magnitude $\kappa$, a microscopic quantity, which is thus finite even
as ${\omega_c\rightarrow 0}$.
Writing that the total quartic term is a sum of the microscopic ones, leads to
${E_4\sim N_{\text{loc}} \kappa \delta_{\text{loc}}^4}$,
implying that ${\chi \sim \kappa/N_{\text{loc}}}$.
This scaling relation is confirmed empirically for QLMs in \cref{sec-quartic-prefactor}.
We thus obtain ${\delta_{\text{loc}}^2 \sim m\omega_c^2/\kappa}$ and
${E_{\text{loc}}\sim m^2 \omega_c^4 N_{\text{loc}}/\kappa}$.
In summary, we get the following scaling description (disregarding constant prefactors):
\begin{equation}
    \label{eq:scaling}
    N_{\text{loc}}\sim \frac{1}{\omega_c},
    \quad
    E_{\text{loc}}\sim \omega_c^3 ,
    \quad
    \delta_{\text{loc}}\sim\omega_c,
    \quad
    \ell_{\text{loc}}\sim \frac{1}{\sqrt \omega_c}
    \, .
\end{equation}
Thus, we predict that close to an instability
(\textit{e.g.}~${\omega_c \sim \sqrt{\epsilon}\sim \sqrt{T_c-T}}$),
hopping processes are extended with small characteristic displacement and energy scales.
Away from an instability, on the contrary, they become localized with large displacements and
energy.

\section{Gapped glasses}
\label{sec-gapped-glasses}

We first test \cref{eq:scaling} in three-dimensional gapped glasses,
obtained with ‘breathing’ particles \cite{Brito18,Kapteijns19}.
We use the protocol and parameters of \cite{Wencheng20} reviewed in \cref{sec:parameters}.
In a nutshell,
we perform molecular dynamics (MD) simulations in which the radius of all $N$ particles is an
additional degree of freedom,
whose stiffness $K$ controls the particle polydispersity.
A long run at finite temperature is followed by an instantaneous quench using the ``FIRE''
algorithm to zero temperature.
We then freeze the particle radii and measure the vibrational spectrum,
which presents a gap of magnitude $\omega_c$ that strongly depends on $K$.

Next, we study elementary excitations using thermally activated rearrangements.
They are obtained by heating our samples with standard (non-breathing) MD to a temperature $T_a$
for a duration $t_a$,
followed by an instantaneous quench using again the ``FIRE'' algorithm to zero temperature.
In practice, $T_a$ and $t_a$ are chosen so as to trigger one rearrangement per sample in average.
In practice we observe up to 4 rearrangements per sample in practice,
which we then separate in individual ones using an algorithm developed in \cite{Wencheng20}.
A displacement field $|\delta \bm{R}\rangle\equiv \{\delta \bm{R}_i\}, {i=1 \ldots N}$,
which is a vector of dimension $Nd$,
is associated to each excitation.
We focus on elementary excitations that go to higher energy states.
In doing so, we eliminate events where one double well is very asymmetric and lies close to a
saddle-node bifurcation
(and would then present a tiny activation barrier not captured by our scaling assuming a symmetric
well).
These events can also be suppressed if the quench is not instantaneous, as we use below for regular
glasses.

For each $\omega_c$, we obtain of the order of $100$ excitations.
We consider the median of the following observables:
\textit{(i)}~The number of particles involved in a rearrangement
${N_\text{loc} \equiv N P_r}$,
where
$P_r \equiv \sum_{i} \left( \delta\bm{R}_{i} \right)^{2} /
    \left( N\sum_{i} \left( \delta\bm{R}_{i} \right)^{4} \right)$
is the participation ratio of $|\delta \bm{R}\rangle$.
\textit{(ii)}~The particle characteristic displacement
${\delta_\text{loc} \equiv X / \sqrt{N_\text{loc}}}$
where $X = ||\delta \bm{R} ||$.
\textit{(iii)}~The length $\ell_\text{loc}$,
defined from the second moment of the position of the particles involved in the rearrangement.
Namely,
${\ell_\text{loc} \equiv 2 \sqrt{I}}$
where $I \equiv \sum_i m_i || \Delta {\bf R}_i ||^2$,
$m_i = || \delta {\bf R}_i ||^2 / \sum_i || \delta{ {\bf R}_i} ||^2$
and
$\Delta {\bf R}_i= {\bf R}_i-\sum_j m_j {\bf R}_j$ is the relative position of particle $i$ with
respect to the center of the rearrangement.
\textit{(iv)}~The energy difference before and after the rearrangement $E_\text{loc}$.
We compare this last quantity to another estimate of the characteristic energy,
obtained in \cite{Wencheng20}.
In particular, after thermal cycling, the density of quasi-localized modes no longer presents a
gap.
Rather it presents a pseudo-gap ${D_L(\omega)=A_4 \omega^4}$ for small $\omega$.
From it, we extract a characteristic energy scale by fitting $A_4(T_a)$ by an Arrhenius behavior.

\begin{figure}[hbt!]
    \centering
    \includegraphics[width=\linewidth]{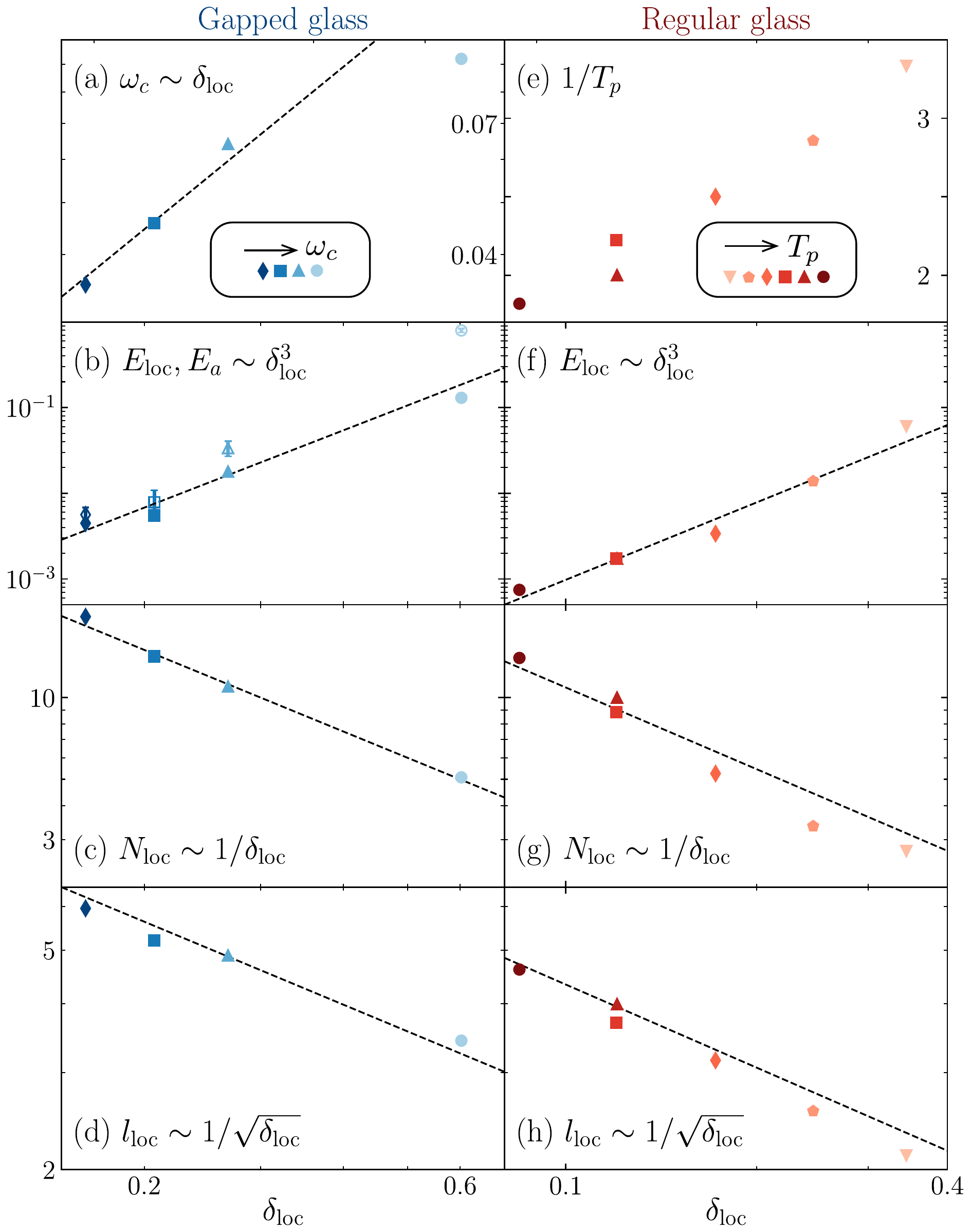}
    \caption{
        Test of our scaling predictions in gapped (left) and regular (right) glasses.
        As a function of the particle characteristic displacement $\delta_\text{loc}$,
        from top to bottom:
        (a)~Gap magnitude $\omega_c$ for the gapped glasses (different markers and color),
        and
        (e)~inverse of the parent temperature $T_p$ for regular glasses,
        at different reheating temperature $T_a$ (different color and markers).
        (b, f)~Energy difference of a local excitation $E_\text{loc}$, and (b)
        for gapped glasses also its global proxy $E_a$ obtained from the density of
        quasi-localized modes \cite{Wencheng20} (open markers).
        (c, g)~Number of particles $N_\text{loc}$ in a rearrangement.
        (d, h)~Characteristic length $\ell_\text{loc}$ of a rearrangement.
        Note that we report the frequency $\omega$ in units of the Debye frequency $\omega_D$,
        the parent temperature $T_p$ in units of the microscopic energy divided by
        the Boltzmann constant,
        the energy $E$ in units of the system's interaction energy density $u_0$,
        and length ($\ell$ and $\delta$) in terms of the typical inter-particle distance $d_0$;
        details in \cref{sec:parameters}.
    }
    \label{fig:scaling}
\end{figure}

Our results are presented in \cref{fig:scaling} (left column) (see \cref{sec-pdf-hopping-processes}
for the whole distributions):
the vanishing scale of particle displacement ${\delta_{\text{loc}}\sim \omega_c}$ is tested in
panel (a),
and ${E_\text{loc}\sim \delta_\text{loc}^3}$ in (b).
It is found to be slightly smaller but comparable to the previously reported quantity $E_a$ (open
markers) \cite{Wencheng20}.
${N_\text{loc}\sim 1/\delta_\text{loc}}$ is tested in panel (c),
and
${\ell_\text{loc}\sim 1/\sqrt{\delta_\text{loc}}}$ in (d).
Overall, we find a good agreement between our scaling predictions and measurements.

\section{Regular ultrastable glasses}
\label{sec-regular-glasses}

In regular glasses,
the density of QLMs does not display a gap \cite{Baity15,Lerner16, Mizuno17} even before reheating.
Indeed, at any finite temperature a gap must necessarily fill up in finite dimensions, because some
excitations transition to their high energy state.
If the latter is barely stable, a low-frequency quasi-localized mode appears (see illustration in
\cref{sec-filling-gap-finite-temp-finite-dim}).
This effect does not affect the scalings of excitations (which are unchanged), but it leads to the
emergence of a pseudo-gap \cite{Wencheng20} that
makes the characteristic frequency $\omega_c$ hard to extract.

Moreover, in finite dimension due to the spatial heterogeneity of the material, the distance to an
elastic instability must vary spatially.
This effect is a relevant perturbation: exponents entering \cref{2}
should depart from their mean-field value \cite{biroli2018random,franz2011field}.
Yet, if these structural fluctuations are small, they will affect exponents only at large length
scales inaccessible in glasses \footnote{The marginality condition $\omega_c=0$ is not observed
    even when quenching from high temperatures, leading to a finite length scale \cite{DeGiuli14}.
}, and mean-field exponents will be observed.
As shown above, it appears to be the case in our very homogeneous gapped glass.
Yet, it may not be so in `normally` prepared ultra-stable glasses.

To study this question, we focus on the local scaling relationships that follow from
\cref{eq:scaling}, which can readily be tested.
Configurations equilibrated by swap at different parent temperature $T_p$ are taken from
\cite{Rainone20} who used the specific liquid model of \cite{Lerner19}, that we then
instantaneously quench
from $T_p$ to ${T_p/3}$, followed by a small cooling rate $\dot{T}$ to zero temperature.
This preparation protocol is used so as to minimize the number of modes that are close to an
instability.
Thus it limits the number of excitations that go to a lower energy state upon temperature cycling,
and allows us to more easily sample statistics
on the excitations increasing energy.

For each $T_p$, we again obtain of the order of $100$ excitations of these inherent structures
through temperature cycles at a low $T_a$ and a short time $t_a$, from which
$N_{\text{loc}}$, $E_{\text{loc}}$, $\ell_{\text{loc}}$ and $\delta_{\text{loc}}$ are then
extracted.
As shown in \cref{fig:scaling}, right column, we again find a very good agreement with our
predictions:
the geometrical description of localized excitations that follows from \cref{eq:scaling} appears to
hold also in regular glasses.
We checked that the scaling predictions also hold well at other two higher reheating temperatures
$T_a$ (see \cref{sec-test-scaling-regular-glasses-reheating}).
Thus a simple mean-field approach already captures the geometry and energy of excitations quite
satisfyingly \footnote{When the displacements $\delta$ becomes of order of the particle size, we
    found for the gapped glass that a small fraction of excitations are strings \cite{Wencheng20}.
    As we will report elsewhere, it is also true for regular glasses.
    Although strings are presumably not accurately described by mean-field argument,
    their fraction is small: removing them from the statistics does not change our observations.
}.

Our second claim is that the increased stability upon cooling predicted by mean-field methods
(corresponding to a growing characteristic frequency ${\omega_{c}(T_p)}$ as $T_p$ decreases),
together with our scaling relations \cref{eq:scaling},
imply that in regular glasses local excitations must then become less extended and involve fewer
particles
--precisely as has been observed in the literature
\cite{Lerner16,Mizuno17,Lerner18,Scalliet19,Wang19,Khomenko20,Rainone20}, and confirmed in
\cref{fig:scaling}(g,h).
We further predict that the characteristic energy of excitations and the displacement should
(rapidly) increase upon cooling (\textit{i.e.}\ decreasing $T_p$)
as confirmed in \cref{fig:scaling}(e,f).

Note that a (crude) estimate of some effective ${\omega_{c}(T_p)}$ can be obtained by comparing
the displacement magnitude $\delta_{\text{loc}}$ of the lowest-energy excitations in our samples,
to those of gapped samples, \textit{i.e.}~comparing Figs.~\ref{fig:scaling}(a) and~(e), which
corresponds to a rapidly growing characteristic frequency upon cooling (see \cref{sec-omega_c} for
details).

\section{Conclusion}
\label{sec-conclusion}

We have developed a scaling description for the architecture of local excitations in glasses,
expressed in terms of the distance to an elastic instability where their characteristic length
diverges.
In gapped glasses obtained with breathing particles, this distance is embodied in the magnitude of
the gap $\omega_c$.
This description appears to provide guidance in regular glasses as well,
where a characteristic frequency is more challenging to identify from the vibrational spectrum
\cite{Wencheng20}.

Using the mean-field result that the gap $\omega_c$ grows upon cooling together with our arguments
explains why excitations become less extended upon cooling,
and leads to two other confirmed predictions.
First, excitations have larger displacements in stable glasses.
Second, we predict a rapidly growing low energy scale for local excitations,
corresponding to two decades in the temperature range probed as apparent in \cref{fig:scaling}(f).
The density of two-level systems should be diminished by this growing energy,
as observed numerically \cite{Khomenko20,Khomenko21},
since it implies a larger tunneling barrier that will eventually become hard to overcome by quantum
fluctuations on experimental time scales \cite{Wencheng20}.

Note that our mean-field arguments appear to yield appropriate exponents in three-dimensional
simulations, at least in the limited range accessible in glasses
\cite{berthier_biroli_2011_RevModPhys83_587}.
This situation is reminiscent of the jamming literature
\cite{DeGiuli14,book_parisi_urbani_zamponi_2020},
and suggests that structural disorder in glasses induces limited heterogeneities in their elastic
properties.
It would be interesting to design a Ginzburg criterion, in the spirit of
\cite{franz_2012_PNAS109_18725},
to estimate beyond which length scale finite dimensional effects could be detectable.

Another interesting question concerns the differences between the geometry of QLMs and excitations
\cite{Khomenko21}.
In particular, the length scale $\ell_c$
below which continuum elasticity breaks down when a force dipole is exerted \cite{Lerner14},
reported to characterize the core of QLMs \cite{Shimada18,Rainone20}, also decreases under cooling
\cite{Rainone20}.
In \cref{sec-measurement-lc}, we observe that $\ell_c$ indeed decouples from the excitations length
scale $\ell_{\text{loc}}$ in our most stable glasses (gapped or regular).

Looking forward, the scaling description of low-energy elementary excitations in glasses may give a
new handle to describe hopping processes in glasses, going beyond simple `elastic' models proposed
in the past \cite{dyre2006colloquium}.
A positive item is their predicted rapidly growing energy scale $E_\text{loc}$ under cooling,
reminiscent of the fragility of liquids.
Yet, a description of all elementary excitations (not only the low-energy ones studied here) is
ultimately needed to make progress on that long-standing question.

\begin{acknowledgments}

    We thank M.~M{\"u}ller and M.~Popovi{\'c} for discussions at the earlier stages of this work,
    and G.
    Biroli, J.P.~Bouchaud, S.~Franz and D.~Reichman for discussions.
    We thank the authors of \cite{Rainone20} for providing us with equilibrated swap configurations
    and discussions.
    T.G.~acknowledges support from the Swiss National Science Foundation (SNSF)
    by the SNSF Ambizione Grant PZ00P2{\_}185843,
    E.A.
    by the SNSF Ambizione Grant PZ00P2{\_}173962.
    We thank the Simons Foundation Grant (\#454953 Matthieu Wyart) and the SNSF under Grant
    No.~200021-165509 for support.

\end{acknowledgments}

\appendix

\section{Derivation of correlation function~\texorpdfstring{$G$}{G}}
\label{sec-derivation-correlation-function}

For pedagogical completeness, here we rederive the correlation function $G(r-r')$ of the overlap
$Q(r)$, which is a standard result from the Ginzburg-Landau theory developed in critical phenomena.
Note that we denote the distance $r - r'$ explicitly,
while in the main text we use translational invariance to render the notation shorter.

Let us define $\phi(r) \equiv Q(r) - Q_{eq}$.
The two-point correlation function is defined as an average over all possible overlap
configurations, with a weight given by the free-energy ${F[\phi]}$.
\begin{widetext}
    It thus reads:
    \begin{align}
        G(r-r')
        &= \left\langle \phi(r) \phi(r') \right\rangle \nonumber
        \\
        &= \int \mathcal{D}_\phi\, \phi(r) \phi(r') \exp(-\beta F[\phi]) \nonumber
        \\
        &= \int \mathcal{D}_\phi\, \phi(r) \phi(r')
        \exp\left\lbrace-\beta \int d^{d}r \left[
            2\sqrt{\epsilon} (\phi(r))^{2} + \tfrac{1}{2}\left(\nabla\phi(r)\right)^{2}
            \right]\right\rbrace\nonumber
        \\
        &= \int \mathcal{D}_\phi\, \phi(r) \, \phi(r')
        \exp\left\lbrace - \int d^{d}r\int d^{d}r' \left[
        \phi(r') \underbrace{
            \beta \delta^{(d)}(r-r') \left(2\sqrt{\epsilon} - \tfrac{1}{2}\nabla^{2}\right)
        }_{
        \sim G^{-1}(r-r')
        }
        \phi(r)
        \right]\right\rbrace.
    \end{align}
\end{widetext}
Note that $\int \mathcal{D}_\phi$ is a functional integral over all possible overlap
configurations, and the last line is obtained by integrating by parts.

Because this path integral has been put in a quadratic form, this simply amounts to computing a
Gaussian integral where the operator $\delta^{(d)}(r-r')\left( 2 \sqrt{\epsilon} - \tfrac{1}{2}
    \nabla^2 \right) \sim G^{-1}(r-r')$ is the functional inverse of the correlator ${G(r-r')}$.
This literally means that
$G^{-1}(r-r')$ and $G(r-r')$ must satisfy the relation
\begin{equation}
    \begin{split}
        & \int d^{d}r'G^{-1}(r-r')G(r'-r'')=\delta^{(d)}(r-r'')
        \\
        & \Rightarrow \quad
        \left(2\sqrt{\epsilon} - \tfrac{1}{2}\nabla^2 \right)G(r-r') \sim\delta^{(d)}(r-r').
    \end{split}
\end{equation}
This differential equation can be easily solved in Fourier space, for instance.
In direct space, when $\left|r-r'\right|/\xi<1$, we reach the Eq.~(3) in the main text:
\begin{align*}
    G(r-r') & \sim\frac{1}{\left|r-r'\right|^{d-2}}\exp\left(-|r-r'|/\xi\right)
\end{align*}
with ${\xi=\frac{1}{2}\epsilon^{-1/4}}$.

\section{Parameters}
\label{sec:parameters}

\begin{table*}[!th]
    \caption{
        Parameters beyond those listed in \cite{Wencheng20,Lerner19, Rainone20}.
    }
    \label{tab:parameters}
    \begin{centering}
        \begin{tabular}{|c||c|c|c|c|c|c|c|c|c|c|}
            \hline
                                 & \multicolumn{10}{c|}{Gapped glasses}\tabularnewline
            \hline
            $\tilde{\omega}_{c}$ & $\omega_{D}$                                          & $u_{0}$  & $d_{0}$ & $T_{a}$                 & $t_{a}$ & $n_{\textrm{all}}$ & $n_{\textrm{pos}}$ & separation & $G$      & $B$\tabularnewline
            \hline
            $\text{1.64}$        & $17.794$                                              & $2.0643$ & $0.738$ & $0.15$                  & $500$   & $655$              & $442$              & yes        & $21.488$ & $78.591$\tabularnewline
            \hline
            $1.19$               & $18.686$                                              & $4.4520$ & $0.918$ & $0.07$                  & $500$   & $915$              & $\,175$            & yes        & $18.570$ & $73.843$\tabularnewline
            \hline
            $0.85$               & $18.698$                                              & $5.3497$ & $0.963$ & $0.03$                  & $500$   & $1245$             & $117$              & yes        & $17.581$ & $72.993$\tabularnewline
            \hline
            $0.65$               & $18.565$                                              & $5.8704$ & $0.988$ & $0.01$                  & $500$   & $1823$             & $95$               & yes        & $16.840$ & $72.599$\tabularnewline
            \hline
            \multicolumn{11}{|c|}{}\tabularnewline
            \hline
                                 & \multicolumn{10}{c|}{Regular glasses} \tabularnewline
            \hline
            $T_{p}$              & $\omega_{D}$                                          & $u_{0}$  & $d_{0}$ & $T_{a}$                 & $t_{a}$ & $n_{\textrm{all}}$ & $n_{\textrm{pos}}$ & separation & $G$      & $B$\tabularnewline
            \hline
            $0.30$               & $18.134$                                              & $4.8245$ & $1.305$ & $\{0.4,0.5,0.6\}$       & $100$   & $\{115,196,406\}$  & $\{95,169,360\}$   & no         & $14.267$ & $44.032$\tabularnewline
            \hline
            $0.35$               & $17.718$                                              & $4.9115$ & $1.305$ & $\{0.1,0.2,0.3\}$       & $100$   & $\{197,328,432\}$  & $\{135,235,324\}$  & no         & $13.592$ & $44.542$\tabularnewline
            \hline
            $0.40$               & $17.298$                                              & $4.9870$ & $1.305$ & $\{0.01,0.02,0.05\}$    & $100$   & $\{167,246,406\}$  & $\{80,127,260\}$   & no         & $12.930$ & $44.982$\tabularnewline
            \hline
            $0.45$               & $16.841$                                              & $5.0551$ & $1.305$ & $\{0.005,0.01,0.02\}$   & $100$   & $\{283,417,532\}$  & $\{96,179,264\}$   & no         & $12.233$ & $45.374$\tabularnewline
            \hline
            $0.50$               & $16.361$                                              & $5.1147$ & $1.305$ & $\{0.003,0.005,0.01\}$  & $100$   & $\{411,560,794\}$  & $\{121,181,294\}$  & no         & $11.523$ & $45.726$\tabularnewline
            \hline
            $0.55$               & $15.928$                                              & $5.1668$ & $1.305$ & $\{0.001,0.002,0.005\}$ & $100$   & $\{280,450,798\}$  & $\{69,107,264\}$   & no         & $10.907$ & $46.028$\tabularnewline
            \hline
        \end{tabular}\\
        \par\end{centering}
\end{table*}

We list all parameters beyond those listed in \cite{Wencheng20,Lerner19,Rainone20}.
For gapped glasses we use ensembles comprising
${n = 10^3}$ samples at ${N = 8000}$ particles in three dimensions for four different gap
frequencies $\tilde{\omega}_c$
(${\tilde{\omega}_c\equiv\omega_c\omega_D}$, see below for the definition of $\omega_D$),
prepared by \cite{Wencheng20}.
For regular glasses we use ensembles comprising
${n = 10^4}$ configurations at ${N = 2000}$ particles in three dimensions for six different parent
temperatures $T_p$,
prepared by \cite{Rainone20}.
The relevant parameters are listed in \cref{tab:parameters},
where, in addition to the parameters described in the text:
\begin{itemize}

    \item $d_0$, the typical inter-particle distance, is defined as the peak in the
          particle-particle correlation function.

    \item $n_\textrm{all}$ is the total number of excitations triggered using temperature cycling;
          $n_\textrm{pos}$ is the number of excitations going to higher energy minima
          (`positive' excitations).

    \item ${\omega_D = \left[18 \pi^2 \rho / \left( 2 c_t^{-3} + c_l^{-3} \right)\right]^{1/3}}$
          is the Debye frequency,
          with the particle number density $\rho \equiv N / V$ and $V$ the volume;
          $c_t = \sqrt{G / (m \rho)}$
          and
          $c_l = \sqrt{(B + 4 G/3 )/ (m \rho)}$
          are the transverse and longitudinal velocity,
          related to the shear modulus $G$ and bulk modulus $B$;
          $m$ is the particle mass (taken equal for all particles).

    \item $u_0$ is the summation of pair interaction energy divided by $N$.

\end{itemize}
Note that $\omega_D$, $G$, $B$, and $u_0$ are obtained as average values of sample-to-sample
fluctuating quantities.

The units in \cref{tab:parameters} are as follows.
Length ($d_0$) is in units of $D_0$,
the initial diameter of small particles in gapped glasses,
and the diameter of smallest particles in regular glasses
(particles sizes are inverse power law distributed, and $D_0$ is the smallest diameter
\textit{i.e.}~the lower bound of the diameter distribution).
Energy ($u_0$) is in units of $\epsilon_0$, the prefactor of the inter-particle interaction
potential.
Temperature ($T_a$ and $T_p$) is in units of $\epsilon_0 / k_B$, where we set Boltzmann's constant
$k_B$ to $1$.
Time ($t_a$, $\tilde{\omega}_c^{-1}$, $\omega_D^{-1}$) is in units of $\sqrt{m D_0^2/\epsilon_0}$,
where $m$ is the particle mass (equal for all particles).
Bulk modulus $B$ and shear modulus $G$ are in units of $\epsilon_0 / D_0^3$.

Note, furthermore, that:
\textit{(i)}~In preparing regular glasses,
a protocol is adopted where we instantaneously quench to $T_p / 3$,
and then slowly quench rate at a rate $\dot{T} = 10^{-3}$
so that the fraction of `positive' excitation is not low (see \cref{tab:parameters}).
We checked that if the glasses are instead prepared by instantaneous quench
(like we do for the `breathing' particles),
not more than $5\%$ of excitations are `positive' excitations at the highest $T_p$ we consider,
which is inefficient to obtain good statistics.
\textit{(ii)}~Since less than 10\% of samples rearrange in regular glasses,
we assume that each rearrangement is an elementary excitation,
and we do not apply our separation algorithm \cite{Wencheng20}.
\textit{(iii)}~For the `breathing particles' the pressure is fixed to a constant value.
This is why $d_0$ is different at different $\omega_c$,
and we adopt the notation $V \equiv \langle V_s \rangle$, with $V_s$ the volume of the individual
samples.
In regular glasses, instead, the volume is fixed to a constant value.

\section{Quartic term for quasi-localized modes}
\label{sec-quartic-prefactor}

\begin{figure}[!htp]
    \centering
    \includegraphics[width=\linewidth]{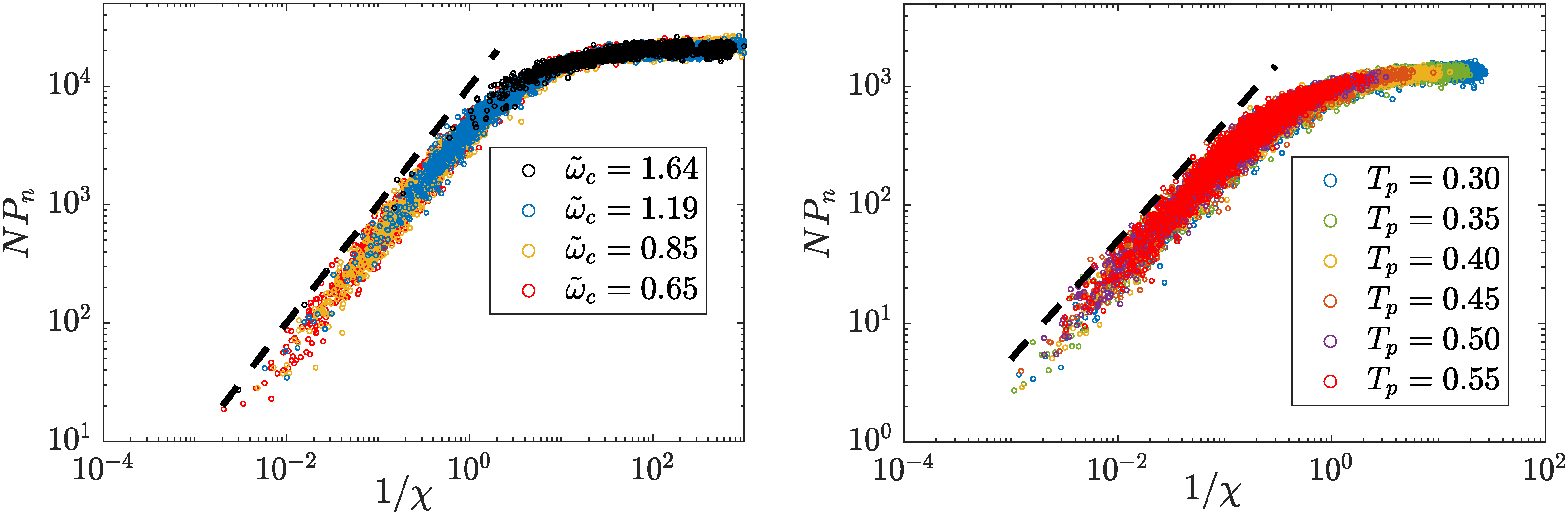}
    \caption{
        $NP_n$ vs $1/\chi$ in gapped glasses (left) and in regular glasses (right).
    }
    \label{fig:chi}
\end{figure}

Here we show that the participation ratio $P_n$ of quasi-localized modes
is proportional to the inverse of the coefficient $\chi$ of quartic term along quasi-localized
modes,
both in gapped glasses and in regular glasses.
In particular, $N P_n\equiv 1 / \sum_{i} || \boldsymbol{\Psi}_{i} ||^4$
where $i$ is the eigenmode component on the $i$th particle.
$\chi \equiv (\partial_{\boldsymbol{r}_\alpha}^4 U)_{ijkl} \boldsymbol{\Psi}_i \boldsymbol{\Psi}_j \boldsymbol{\Psi}_k \boldsymbol{\Psi}_l$
where the $(\partial_{\boldsymbol{r}_\alpha}^4 U)_{ijkl}$ is the fourth order (spatial) derivative of the total interaction potential energy,
it is a rank four tensor of size $(Nd)^4$.
For details see \cite{Gartner16,Wencheng20}.
The scatter plots in \cref{fig:chi}, show that at low $NP_n$ the scaling is consistent with
$NP_n\sim 1/\chi$.
We use 25 samples at $N=32000$ gapped glasses and 100 samples at $N = 2000$ in regular glasses to
calculate $NP_n$ and $\chi$,
and these samples are also used in \cref{sec-measurement-lc}.

\section{Distributions for hopping processes}
\label{sec-pdf-hopping-processes}

\begin{figure*}[!htp]
    \begin{centering}
        \includegraphics[width=\linewidth]{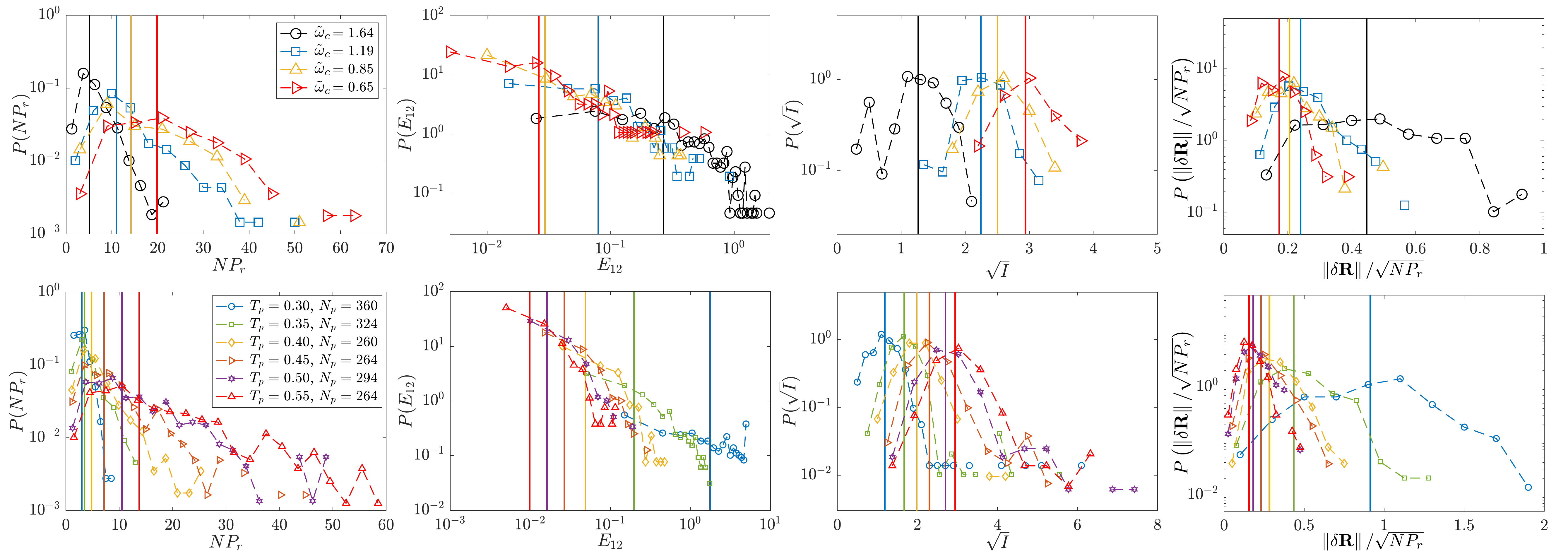}
        \par\end{centering}
    \caption{
        The first row shows the corresponding distributions for gapped glasses.
        The second row shows the distributions for regular ultrastable glasses at the highest $T_a$ that we
        consider (see \cref{fig:threeTa}), as it corresponds to the highest number of excitations.
        The solid vertical lines indicate the median values
        $N_{\text{loc}}$,
        $E_{\text{loc}}u_0$,
        $\ell_{\text{loc}}d_0/2$,
        and $\delta_{\text{loc}}d_0$
        that are discussed in main text.
    }
    \label{fig:distribution}
\end{figure*}

\cref{fig:distribution} shows the distributions of the bare quantities
$NP_r$,
of the energy difference $E_{12}$,
of the length $\sqrt{I}$,
and $|| \delta \boldsymbol{R}|| / \sqrt{NP_r}$
of thermally-activated rearrangements, \textit{i.e.}\ `hopping processes'.
The medians are shown using a vertical line,
defining $N_\text{loc} \equiv \text{median} (N P_r)$,
$E_\text{loc} \equiv \text{median} (E_{12} / u_0)$,
$\ell_{\text{loc}} \equiv \text{median} (2 \sqrt{I} / d_0)$,
and $\delta_\text{loc} \equiv \text{median} (|| \delta \boldsymbol{R}||) / (N_\text{loc}d_0)$.
Data are taken with the conditioning on excitations going to higher energy state:
$E_{12}>0$ both in gapped glasses (first row) and regular glasses (second) row.
Distributions are peaked around a maximum, except for the energy distribution whose distribution is
maximum in zero
(corresponding to symmetric double wells).

\section{Filling up the gap}
\label{sec-filling-gap-finite-temp-finite-dim}

\begin{figure}[!htp]
    \centering
    \includegraphics[width=.95\linewidth]{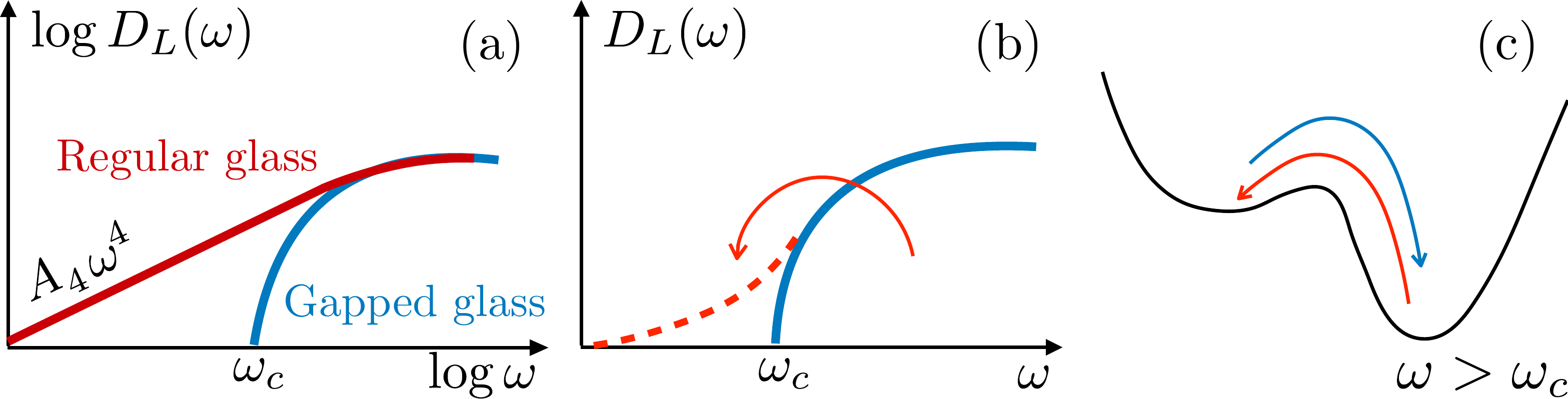}
    \caption{
        (a)~Sketch of density of QLMs in log-log scale for a ``gapped'' (blue) and
        a ``regular'' (red) glass.
        (b)~Sketch of the reservoir picture \cite{Wencheng20},
        in which a gap in the vibrational spectrum is filled-up by thermally activating
        excitations of frequencies $\omega>\omega_c$.
        This process (c) corresponds to excitations
        moving from a lower energy minima with a characteristic frequency $\omega\approx \omega_c$
        to a higher one with characteristic frequency $\omega<\omega_c$.
    }
    \label{fig:sketch_2}
\end{figure}

\cref{fig:sketch_2} shows that, in finite dimension,
the gap in the vibrational spectrum is filled up by thermally activating excitations.
In particular, we expect that upon reheating with a small temperature,
the vibrational spectrum is filled up with modes with excitations from a `reservoir' of
excitations at frequencies $\omega>\omega_c$ \cite{Wencheng20}.
This process corresponds to excitations
moving from a lower energy minima with a characteristic frequency $\omega\approx \omega_c$ to
a higher one with characteristic frequency $\omega<\omega_c$.
Ultimately, this effect leads to a pseudo-gap
$D_L(\omega) =A_4 \omega^4$ at $\omega<\omega_c$.
Yet, the excitations responsible for populating the gap have a characteristic
frequency $\omega\sim \omega_c$ in their lower-energy state,
and their architecture and energy must follow our
predictions for a gap of magnitude $\omega_c$.

\section{Different reheating temperatures~\texorpdfstring{$T_a$}{T\_a} (regular glasses)}
\label{sec-test-scaling-regular-glasses-reheating}

\begin{figure}[!htp]
    \centering
    \includegraphics[width=\linewidth]{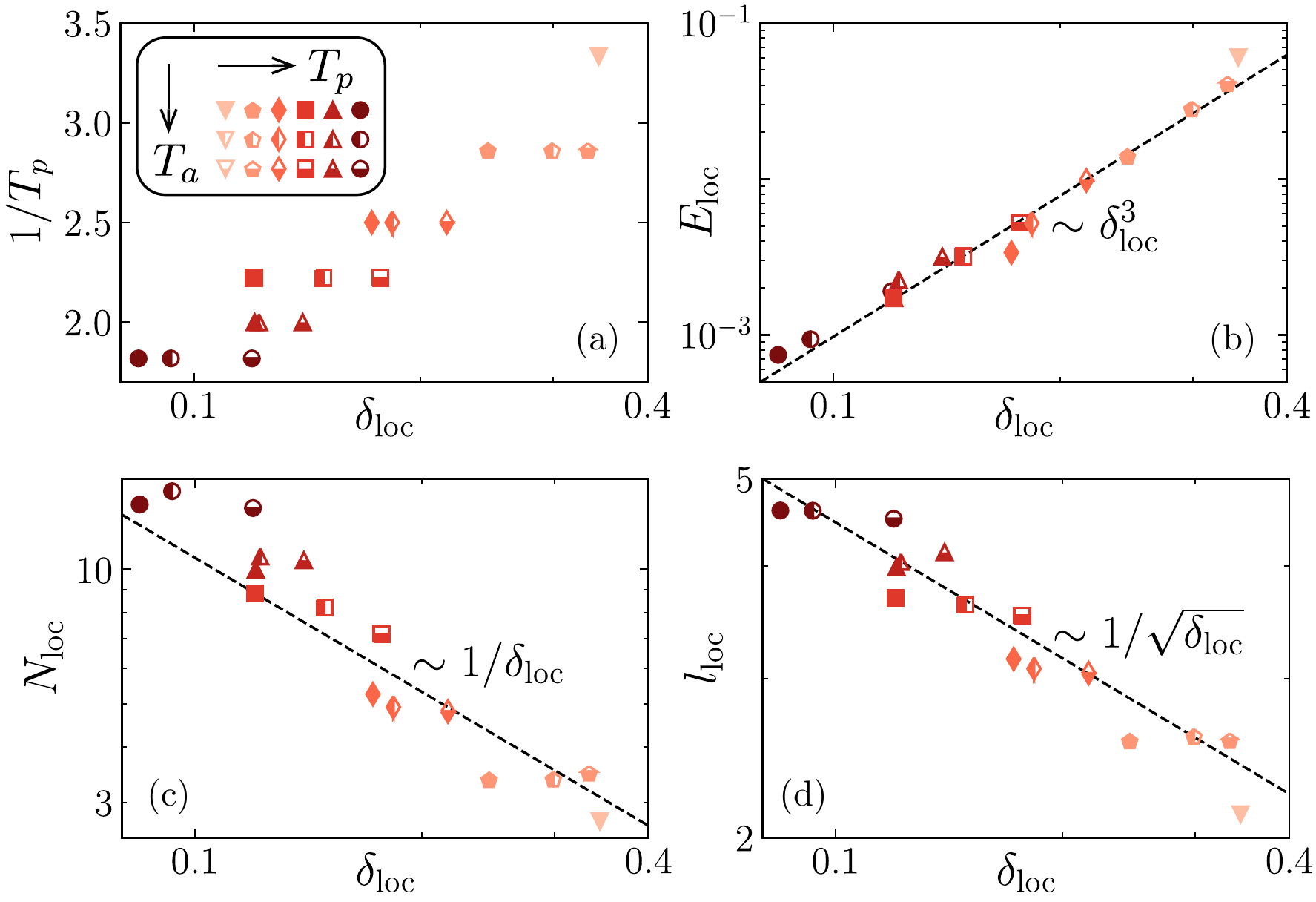}
    \caption{
        Test of our scaling predictions in
        regular glasses for three reheating temperatures $T_a$
        (the reheating duration $t_a$ is kept constant, see \cref{tab:parameters}).
    }
    \label{fig:threeTa}
\end{figure}

\cref{fig:threeTa} investigates the scaling predictions at
three temperature $T_a$ in regular glasses.
The results found are overall
robust toward the change of $T_a$.
We can still discern some systematic effects: excitations have a larger energy, present larger
displacements and involve more particles as $T_a$ increases.

\section{Crude estimate of \texorpdfstring{$\omega_c$}{omega\_c} in regular glass}
\label{sec-omega_c}

\begin{figure}[!htp]
    \centering
    \includegraphics[width=0.5\linewidth]{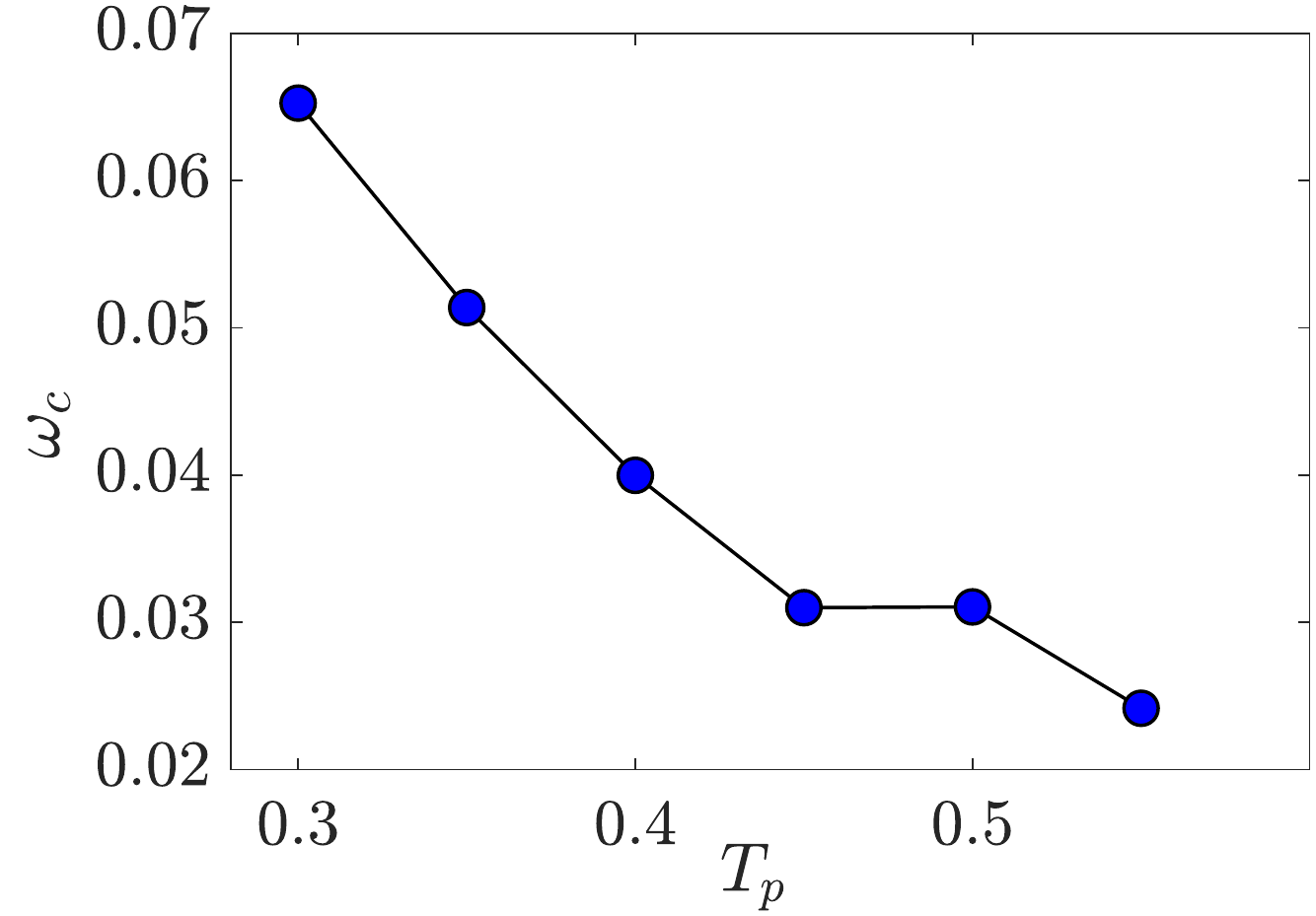}
    \caption{
        The estimated $\omega_c$ \emph{vs} $T_p$, in regular glasses.
    }
    \label{fig:omega_c}
\end{figure}

In this section, we estimate $\omega_c$ in regular glasses.
We suppose in gapped glasses and regular glasses $\omega_c$ varies with $\delta_{\text{loc}}$ in
the same way.
Since we know both $\omega_c$ and $\delta_{\text{loc}}$ in gapped glasses,
we fit the data by $\ln(\omega_c) = c_1+c_2 \ln(\delta_{\text{loc}})$ to extract $c_1$ and $c_2$.
We employ them to get an estimate of $\omega_c$ at the lowest energy excitations
(lowest $\delta_\text{loc}$ at each $T_p$).
The results, in \cref{fig:omega_c}, $\omega_c$ increases with decreasing $T_p$, as we expect.

\section{Measurement of \texorpdfstring{$\ell_c$}{l\_c}}
\label{sec-measurement-lc}

\begin{figure}[!htp]
    \centering
    \includegraphics[width=\linewidth]{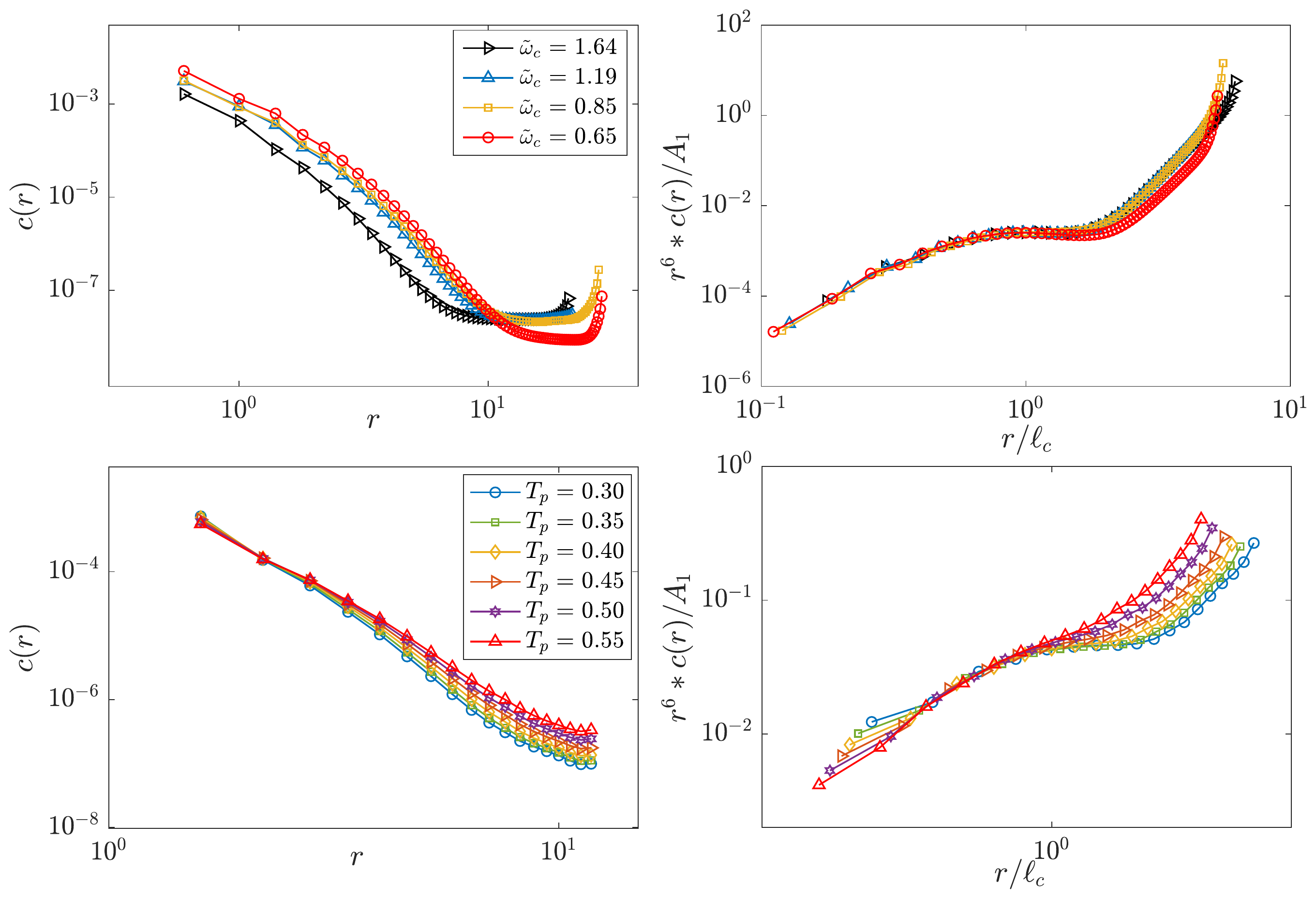}
    \caption{
        Correlation function $c(r)$ of the response to a dipole force response (left) and rescaling
        to extract $\ell_c$ (right),
        for gapped glasses (top) and regular glasses (bottom).
        Note that we find for gapped glasses $A_1=\{1.1,\,6.2,\,10,\,15\}$,
        and for regular glasses $A_1=\{1,\, 1.2,\, 1.4,\, 1.6,\,1.9,\, 2.3\}$.
    }
    \label{fig:lcmeasure}
\end{figure}

To extract the typical global length $\ell_c$,
we perturb the glasses with a local dipole force and look at the correlation function $c(r)$ which
is defined in as \cite{Lerner14,Rainone20}.
$\ell_c$ is defined as the length where rescaling $c(r)$ collapses the data,
see \cref{fig:lcmeasure}.
Note that for this global measurement we use a bigger system
($25$ samples at $N=32000$;
whereby we checked that these $l_c$ collapse the rescaled $c(r)$ at $N=8000$ as well,
except for a small difference at our smallest gap)
at each $\tilde{\omega}_c$ in gapped glasses,
and $100$ samples at $N=2000$ at each $T_p$ in regular glasses to extract $\ell_c$.

\begin{figure}[!htp]
    \centering
    \includegraphics[width=.9\linewidth]{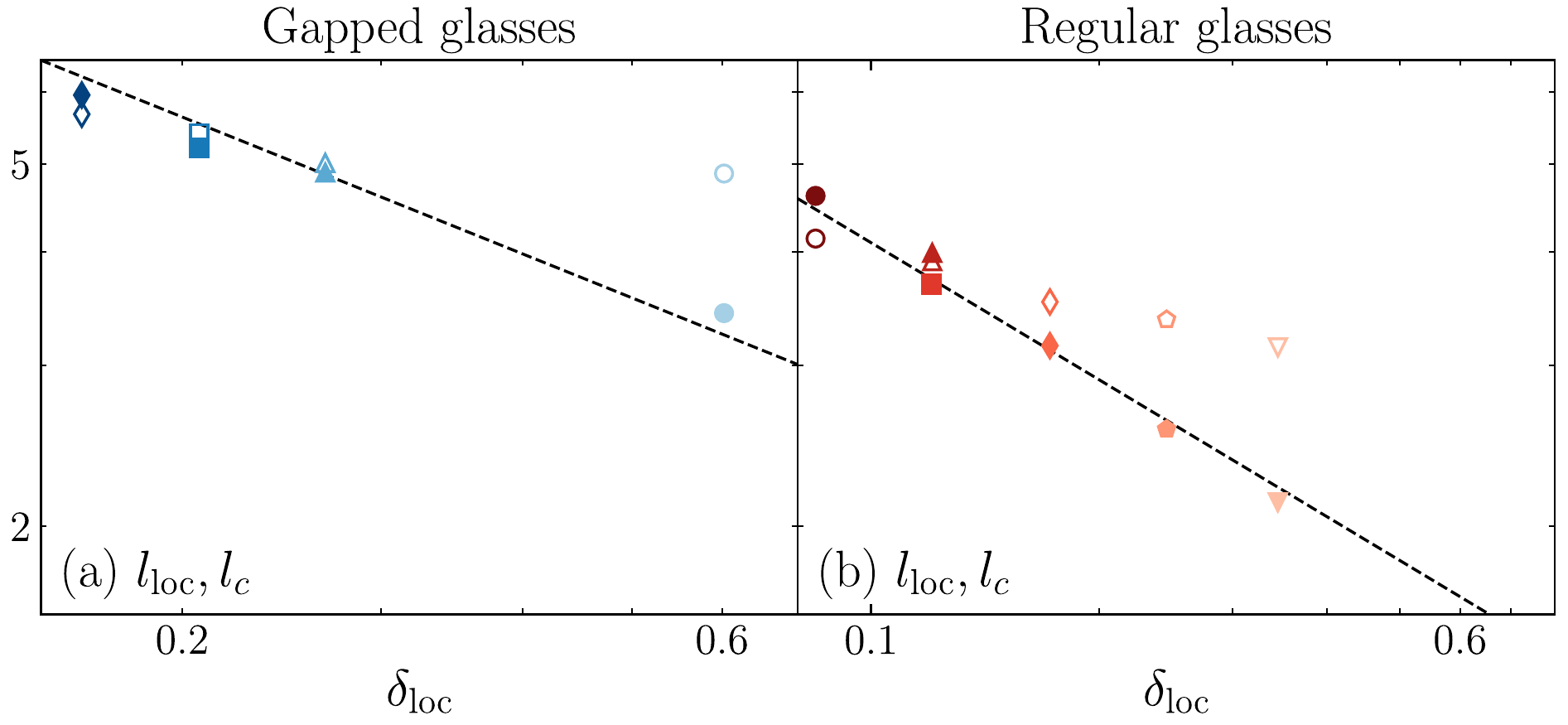}
    \caption{
        Fig.~1(d,h) from the main text with superimposed $\ell_c$ (open markers).
        $\ell_c$ in regular glasses is shown as a function of
        $\delta_\text{loc}$ for the smallest $T_a$ for each parent temperature $T_p$
        (for that reason we only show $\ell_\text{loc}$ for those $T_a$).
        Notice that, like $\ell_{\text{loc}}$, $\ell_c$ is reported in units of $d_0$.
    }
    \label{fig:lc}
\end{figure}

In \cref{fig:lc}, we show that $\ell_c$ decouples from the excitation length scale
$\ell_{\text{loc}}$ in stable (\textit{i.e.}~gapped and regular) glasses,
at large $\delta_{\text{loc}}$.

\clearpage
\bibliographystyle{unsrtnat}
\bibliography{library}
\bibliographystyle{apsrev4-1}

\end{document}